\documentstyle[12pt]{article}
\date{}
\begin{document}

\renewcommand{\theequation}{\thesection.\arabic{equation}}

\title{The $\pi-$Gluon Exchange Interaction Between
Constituent Quarks}

\author{ C. Helminen and D.O. Riska}
\maketitle

\centerline{\it Department of Physics, University of Helsinki,
00014 Finland}

\setcounter{page} {0}
\vspace{1cm}

\centerline{\bf Abstract}
\vspace{0.5cm}

The interaction mediated by irreducible pion and gluon
exchange between constituent quarks is calculated and
shown to have a strong tensor component, which tends
to cancel the pion exchange tensor interaction between
quarks. Its spin-spin component is somewhat weaker than
the pion exchange spin-spin interaction, 
while its central and spin-orbit components
are small in comparison to the corresponding single
gluon exchange interactions.
The combination of the $\pi-$gluon exchange interaction with
the single pion exchange interaction 
and a weak gluon exchange interaction
between constituent quarks has the qualitative features 
required for understanding the hyperfine splittings of
the spectra of the nucleon and the $\Delta$ resonances.\\

\newpage
\centerline{\bf 1. Introduction}
\vspace{0.5cm}

	Both the form and the dynamical origin of the
effective interaction between the constituent quarks,
which form the baryons, remain largely open issues.
The small spin-orbit splittings between the lowest
negative parity states in the baryon spectrum suggest
that both the spin-orbit and tensor components of the
effective quark-quark interaction should be weak. This
rules out a strong quark-gluon coupling, as does the
presence of low lying positive parity states below
the lowest negative parity states in all those sectors of
the baryon spectrum without flavor singlet states.
The latter feature is most readily explained if the
main component of the
hyperfine interaction between the quarks is an
attractive flavor
dependent spin-spin interaction \cite{GloRis}.\\

Such an interaction is an integral component of the
Goldstone boson (i.e. $\pi$,$ K$,$\eta$) exchange
interaction between quarks, and as long as the
associated tensor component is dropped, it is indeed
possible to explain the spectrum of the nucleon and the
strange hyperons up to the small spin-orbit splittings
with that interaction in combination with a linear
confining interaction with conventional strength 
\cite{Pless}.
This however leaves the tensor component of the Goldstone
boson exchange interaction to be explained away, somewhat
as the large
spin-orbit interaction of the gluon exchange interaction
has to be explained away in attempts to describe the
baryon spectrum in terms of gluon exchange alone 
\cite{Carls}.\\

We here point out that if the Goldstone boson exchange
interaction between quarks is complemented with a
fairly weak gluon exchange interaction along with
the associated irreducible $\pi-$gluon exchange interaction
the problem of the tensor component of the former
is ameliorated if not eliminated. This is because
the tensor component of the $\pi-$gluon
exchange interaction is of the same order of magnitude
as the pion exchange interaction, but has the opposite
sign. As a result the net tensor interaction is very
weak. This provides an alternative to balancing
the pion exchange tensor
interaction at short range by a corresponding vector
meson exchange interaction of opposite sign \cite{GloRis}.\\

The $\pi-$gluon exchange interaction has an attractive
spin-spin component, which adds to, but is somewhat weaker than,
that of single pion exchange at short range. 
Its detailed behavior at very short range is very sensitive
to the high momentum behavior of the pion and
gluon exchange interactions. Finally the spin-orbit
and central components of the $\pi-$gluon exchange interaction 
turns out to be very weak. The $\pi-$gluon exchange interaction thus
appears to provide part of the explanation for why the
effective interaction between constituent quarks should
have the form of an attractive flavor dependent 
spin--spin interaction and an at most very
weak tensor interaction.\\

Another issue is to what extent the nonperturbative
vacuum of QCD supports gluon exchange. Cooled lattice
calculations suggests that the quark-gluon coupling
should be very weak \cite{Nege}. The valence-QCD
approximation suggests the presence of a residual
weak but nonzero gluon exchange interaction between
quarks \cite{KFLiu}. We shall here assume that
the gluons either decouple from the constituent quarks below
the confinement scale $\Lambda_{QCD}$ or the chiral 
restoration scale $\Lambda_\chi \simeq $
1 GeV or that their coupling freezes at some small 
value below these momentum scales. This agrees with
recent phenomenological studies of the behavior of
the running QCD coupling strength in the infrared limit
\cite{Matting,Brodsky}.  
In addition we make the conventional assumption
that the Goldstone bosons decouple above the chiral
restoration scale. The $\pi-$gluon exchange
loop mechanisms allow combination of short range
gluon exchange with long range pion exchange.\\

For the calculation of the $\pi-$gluon exchange interaction
we adopt the Blankenbecler-Sugar quasipotential framework,
which allows a covariant extraction of the iterated
single pion and gluon exchange interactions from the
Bethe-Salpeter equation kernel \cite{Blan}. 
The resulting interaction
is therefore real and almost energy independent.
In the calculation we rely largely on the
formalism developed in Refs. \cite{ChuR,Ris}
for a different application.\\

This paper falls into 7 sections. In 
section 2 we define
the $\pi-$gluon exchange interaction. In section 3 we
calculate the $\pi-$gluon exchange loop diagrams. In 
section 4 we derive the iterated pion and gluon
exchange interaction, which has to be subtracted
from the covariant loop diagrams in order to
obtain a properly defined
irreducible interaction. In section 5 the 
adiabatic limit of the $\pi-$gluon exchange
interaction is obtained.
In section 6 we give
numerical results for the interaction components.
Section 7 contains a summarizing discussion.\\

\vspace{1 cm}

\centerline{\bf 2. Definition of the $\pi-$gluon
exchange interaction}

\vspace{0.5cm}

The Bethe-Salpeter equation for the invariant quark-quark
scattering amplitude $M$ may be written as
$$M=K+KGM,\eqno(2.1)$$
where $K$ is the two-quark irreducible interaction kernel and
$G$ is the two quark propagator. The conventional assumption,
which underlies most quark model phenomenology, is that
this may be approximated as a product of free fermion
propagators with effective constituent quark masses. The
Blankenbecler-Sugar reduction amounts to the separation
of this propagator into a quasi-3-dimensional part $\tilde g$
and a residual part $G-\tilde g$ \cite{Blan,Log}. This
separation reduces the Bethe-Salpeter equation (2.1) to
a 3-dimensional quasipotential equation:
$$M=U+U\tilde g M,\eqno(2.2)$$
with an accompanying defining
equation for the quasipotential $U$:
$$U=K +K(G-\tilde g)U.\eqno(2.3)$$
For the 3-dimensional propagator $\tilde g$ we employ the
symmetrical choice \cite{LoPa,CoRis}:
$$\tilde g(\vec k;W)=2\pi i\delta (k_0)
{\Lambda_+ ^{(1)}(\vec k)\Lambda_+ ^{(2)}(-\vec k)
\over E^2(\vec k)-W^2-i\epsilon}.\eqno(2.4)$$
Here $W$ is the total energy and $\Lambda_+^{(i)}(\vec k)$
(i=1,2) are positive energy projection operators
$$\Lambda_+^{(i)}(\vec k)={\gamma_4^{(i)}E(\vec k)-i\vec
\gamma^{(i)}
\cdot \vec k+m_q\over 2m_q},\eqno(2.5)$$
where $m_q$ is the constituent quark mass.\\

Consider now the case, where the irreducible interaction kernel 
$K$ is approximated
as the sum of the one--pion-- and one--gluon--exchange
and the crossed two--pion, two-gluon and
pion+gluon exchange interactions:
$$K\simeq K_\pi+K_G+K_{\pi\pi}+K_{GG}+K_{\pi G}.\eqno(2.6)$$
The lowest (second) order term in the quasipotential (2.3),
which corresponds to the tree approximation is then
the sum of one--pion and one--gluon exchange interactions:
$$U^{(2)}=K_\pi+K_G.\eqno(2.7)$$
The expressions for these 
are readily obtained from the pion- and
gluon-quark coupling Lagrangians:
$${\cal L}_{\pi qq}=i{f_{\pi qq}\over m_\pi}
\bar\psi\gamma_\mu\gamma_5\partial_\mu\vec\phi\cdot \vec \tau\psi,$$
$${\cal L}_{Gqq}=i\sqrt{4\pi\alpha_S}\bar\psi\gamma_\mu \vec G_\mu
\cdot\vec\lambda_C\psi.\eqno(2.8)$$
Here $\vec\lambda_C$ is the vector of $SU(3)$ color
generators and $\vec G_\mu$ is the gluon field. 
For the one-pion and one-gluon exchange interactions
between quarks one then obtains the expressions:
$$K_\pi={f_{\pi qq}^2\over m_\pi^2}
{(\gamma^{(1)}\cdot k)\gamma_5^{(1)}
(\gamma^{(2)}\cdot k)\gamma_5^{(2)}
\over m_\pi^2+k^2}f(k^2)\vec\tau^1\cdot\vec \tau^2,\eqno(2.9a)$$
$$K_G=-{8\pi\over 3}\alpha_S {\gamma^{(1)}\cdot
\gamma^{(2)}\over k^2}g(k^2).\eqno(2.9b)$$
Here $f_{\pi qq}$ is the pion-quark coupling constant, 
and $\alpha_S$ the effective coupling constant for gluons
to constituent quarks. The function $f(k^2)$ is a cut-off 
factor, that should describe the decoupling of the pions
from constituent quarks above the chiral restoration
scale $\Lambda_\chi$, so that $f(k^2=-m^2_\pi)=1$ and
$\lim_{k^2\rightarrow \infty}f(k^2)=0$.\\

We shall here take the cut-off function $f(k^2)$ to have the
form
$$f(k^2)={\Lambda_\chi^2-m_\pi^2\over \Lambda_\chi^2
+k^2},\eqno(2.10)$$
where $\Lambda_\chi$ is the chiral symmetry restoration scale.
The pseudovector pion-quark coupling constant $f_{\pi qq}$
may be related to the $\pi NN$ pseudovector coupling as
$f_{\pi qq}=(3/5)f_{\pi NN} \simeq$ 0.6, a value which
will be adopted here.\\

The function $g(k^2)$ on the other hand is a factor introduced
to describe the momentum dependence of the
quark-gluon coupling constant.
For momenta above $\Lambda_{QCD}$ this factor should
contain the inverse logarithmic fall-off factor
$1/{\rm log}(k^2/\Lambda_{QCD}^2)$. At small values of
momentum transfer it should either vanish
(if gluons decouple completely) or reach a constant
value \cite{Matting,Brodsky}. In the former case
the gluon interaction would be screened at long range,
which would most simply be achieved by taking
$g(k^2)$ to have the form
$$g(k^2)={k^2\over \Lambda^2+k^2}
{{\rm log}({4\Lambda^2\over \Lambda_{QCD}^2})
\over {\rm log}({k^2+4\Lambda^2\over \Lambda_{QCD}^2})}
,\eqno(2.11)$$
where the parameter $\Lambda$, which represents the inverse
of the screening length, is taken to be either $\Lambda_{QCD}$
or $\Lambda_\chi$. This 
in practice implies
that the gluon attains a finite mass in the nonperturbative
regime.
For calculational convenience the expression (2.11) will be
approximated here by the simpler form
$$g(k^2)={k^2\over \Lambda^2+k^2}{1\over 1+
{k^2\over \Lambda_F^2} }.\eqno(2.12)$$
With an appropriately chosen value for $\Lambda_F$ this form
gives a fair approximation
to (2.11) up to $k \simeq$ 5 GeV.
If the screening scale $\Lambda$ is set to $\Lambda_{QCD}$
(0.25 GeV \cite{Don}) and
the effective coupling constant $\alpha_S$ in (2.9b)
is taken to be $\alpha_S=0.42$ and $\Lambda_F$ = 4.3 GeV, the
effective quark gluon coupling at 
the charmonium scale $k=$ 1.3 GeV takes the value
0.37 and at the bottomonium scale $k=$ 4.1 GeV 
the value 0.22 in agreement with the
lattice calculation results values given in Ref. \cite{Lepage}.
At the light constituent quark
scale $k\simeq$ 0.34 GeV the effective
$\alpha_S$ value will then be $\alpha_S\simeq 0.27$,
a value which is small enough that the spectroscopic
problems associated with combinations of single gluon and
pion exchange interactions are avoided \cite{Papp}.
If the screening scale $\Lambda$ in (2.12) is set to
$\Lambda_\chi \simeq$ 1 GeV, the effective $\alpha_S$
value at the light constituent quark mass scale
drops to only 0.04.\\

The fourth order terms in the quasipotential are the
two--pion, two--gluon and the $\pi-$gluon exchange interactions:
$$U_{\pi\pi}=M_{\pi\pi}-K_\pi\tilde g K_\pi,\quad
U_{GG}=M_{GG}-K_G\tilde g K_G,\eqno(2.13a)$$
$$U_{\pi G}=M_{\pi G}-K_\pi\tilde g K_G-K_G\tilde g 
K_\pi.\eqno(2.13b)$$
Here $M_{\pi\pi}$ and $M_{GG}$ are the fourth order
two-pion and two-gluon exchange quark-quark scattering
amplitudes respectively and $M_{\pi G}$ is the fourth
order $\pi-$gluon exchange amplitude.
It is the latter one, and the corresponding quasipotential
$U_{\pi G}$ that we shall consider here, because these
have the same flavor dependence as the one--pion--exchange
interaction $K_\pi$. 
The $\pi-$gluon exchange amplitude is illustrated by the
Feynman diagrams in Fig. 1.
The main part of the two--pion exchange
interaction $K_{\pi\pi}$ is a central interaction 
with weak flavor dependence (cf. \cite{LoPa,Chem}), which
naturally is incorporated into the effective
confining interaction between quarks. Similarly the
two--gluon exchange interaction is flavor independent, and
naturally incorporated into the confining interaction.\\

\vspace{1 cm}

\setcounter{section}{3}
\centerline{\bf 3. The $\pi-$Gluon Exchange Amplitude}

\vspace{0.5cm}

The $\pi-$gluon exchange amplitude will have the same flavor
dependence as the one-pion-exchange interaction (2.9a). Its
spin structure may be expressed in terms of any set of
linearly independent spin invariants. In order to avoid
kinematic singularities, it proves convenient to
employ the following set \cite{ChuR,Ris}:
$$S_1=\gamma_5^{(1)}\gamma_5^{(2)}(\gamma^{(1)}
\cdot\gamma^{(2)}),\quad
S_2=\{(\gamma^{(1)}\cdot N)(\gamma^{(2)}\cdot P),S_1\}$$
$$S_3=i[\gamma^{(1)}\cdot N+\gamma^{(2)}\cdot P]S_1,\quad
S_4=\gamma^{(1)}\cdot\gamma^{(2)} S_1,
\quad S_5=\gamma_5^{(1)}\gamma_5
^{(2)}.\eqno(3.1)$$
Here the 4-vectors $P$ and $N$ have been defined in terms of
the initial $p_1,p_2$ and final $p'_1,p'_2$ quark 4-momenta
as
$$P={1\over 2}(p_1+p'_1),\quad N={1\over 2}(p_2+p'_2).\eqno(3.2)$$
The relation between the invariants $S_j$ and the more common
Fermi invariants $SVTAP$ is given in Ref. \cite{Ris}.\\

The $\pi-$gluon exchange amplitude $M_{\pi G}$ may expressed
in terms of the spin invariants (3.1) as
$$M_{\pi G}=\sum_{j=1}^5s_j(s,t,u)S_j\vec\tau^1\cdot\vec\tau^2.
\eqno(3.3)$$
Here the variables $s,t,u$ are defined as
$$s=-(p_1+p_2)^2,\quad t=-(p'_1-p_1)^2,
\quad u=-(p'_1-p_2)^2.\eqno(3.4)$$

The scalar amplitudes $s_j$ may be expressed in terms of
dispersion relations in $t$ as
$$s_j(s,t,u)={1\over \pi}\int_{t_0}^\infty dt'
{\sigma_j(t';s,u)\over t'-t-i\epsilon}.\eqno(3.5)$$
where $t_0=(m_\pi+m_G)^2$. Here $m_\pi$ is the pion mass and 
$m_G$ is a fictitious gluon mass, which will be set to
zero or to the screening parameter $\Lambda$ (2.11) in the end.
To derive the spectral functions $\sigma(t;s,u)$
we use the unitarity relation in the $t-$channel
for the $\pi-$gluon intermediate states:
$$\sum_j\sigma_j S_j\vec\tau^1\cdot\vec\tau^2\simeq
{|\vec q|\over 32 \pi^2\sqrt{t}}\int d\Omega_q
T^\dagger T,\eqno(3.6)$$
where T is the amplitude for gluoproduction of pions in the
$t-$channel. The diagrams included in the
unitarity integral are those shown in Fig. 1.
Here the variable $|\vec q|$ is defined as
$$|\vec q|={1\over 4t}\big[t^2-2t(m_\pi^2+m_G^2)+
(m_\pi^2-m_G^2)^2\big].\eqno(3.7)$$

The ``Born terms'' in the pion gluoproduction amplitude
$T$ required in the unitarity integral (3.6) are
readily obtained from the coupling Lagrangians (2.8):
with exception of the color factor, the expressions
are
the same as those given in Ref. \cite{Ris} for the
$N\bar N\rightarrow \pi\omega$ amplitude. Note that
the pion gluoproduction amplitude takes the same form for
pseudovector and pseudoscalar
$\pi-$quark coupling.\\
 
These ``Born terms'' in the pion gluoproduction amplitude
then lead to the following expressions for the
$\sigma_j$ in (3.5), which may be obtained by suitable
extension of the formalism in Ref. \cite{Ris}:
\setcounter{equation}{7}
\begin{equation}
\begin{array}{rcl}
\sigma_1(s,t,u)&=&m_q^2\bar N[\theta_1(s,t)-\theta_1(u,t)],\\
&&\\
\sigma_2(s,t,u)&=&-\sigma_1(s,t,u)/2m_q^2,\\
&&\\
\sigma_3(s,t,u)&=&m_q\bar N[\theta_2(s,t)+\theta_2(u,t)],\\
&&\\
\sigma_4(s,t,u)&=&\bar N[\theta_4(s,t)+\theta_4(u,t)],\\
&&\\
\sigma_5(s,t,u)&=&\sigma_5^+(s,t,u)+\sigma_5^-(s,t,u),\\
&&\\
\ \sigma_5^+(s,t,u)&=&-\bar N[4\theta_4(s,t)+{1\over 4}(s-\bar s)
\theta_1(s,t)+(s\leftrightarrow u)],\\
&&\\
\ \sigma_5^-(s,t,u)&=&\bar N\{I_0(s,t)[{1\over 4}(2s-4m_q^2)
(1-\bar\mu^2/t)^2+|\vec q|^2]\\
&&\\
&&+(2m_q^2-{1\over 4}t)\theta_1(s,t)
-{1\over 2}(s-\bar s)\theta_2(s,t)
+{1\over t}(s-\bar s)\theta_4(s,t)-(s\leftrightarrow u)\}.\\
\end{array}
\end{equation}
Here $\bar N$ is defined as
$$\bar N=-{2 f_{\pi qq}^2\alpha_S\over 3\pi}
{|\vec q|\over\sqrt{t}}
({m_q\over m_\pi})^2,
\eqno(3.9)$$
and $\bar s=4m_q^2-t-s$, with $m_q$ being the quark mass.\\

The function $I_0$ in (3.8) is defined as
\setcounter{equation}{9}
\begin{equation}
I_0(x,t)=\left\{ \begin{array}{ll}
 {2\pi\over\sqrt{\xi^2-\eta\zeta}}{\rm log}{\xi+\zeta+
 \sqrt{\xi^2-\eta\zeta}\over\xi+\zeta-\sqrt{\xi^2-\eta\zeta}}, &
 \xi^2-\eta\zeta>0\\
  & \\
 {4\pi\over\xi+\zeta}, & \xi^2-\eta\zeta=0\\
  & \\
 {8\pi\over\sqrt{\zeta\eta-\xi^2}}{\rm arctan}{2q^2x\over
 \sqrt{\zeta\eta-\xi^2}}, & \xi^2-\eta\zeta<0.\\
  \end{array} \right. 
\end{equation}
Here the variables $\eta$, $\xi$, $\zeta$, $\bar\nu$ and
$\chi$ are defined as
$$\eta=4|\vec q|^2x,\qquad \xi=-2|\vec q|^2x,$$
$$\zeta=4|\vec q|^2\chi^2+{1\over 4}(t-\bar\nu^2)^2,
\qquad\bar\nu^2=m_G^2+m_\pi^2,$$
$$\chi^2=m_q^2-{1\over 4}t.\eqno(3.11)$$
The functions $\theta_j$ in (3.8) are defined as

$$\theta_1(x,t)={1\over x\bar x}\Big\{|\vec q|^2(x-\bar x)I_0(x,t)
+{x+\bar x\over\chi^2}[\pi-{1\over 4}x'I(x')]
-x'(2x-\bar x)R(x,t)\Big\},$$
$$\theta_2(x,t)={1\over x\bar x}\Big\{|\vec q|^2(x+\bar x)I_0(x,t)
+{x-\bar x\over\chi^2}
[\pi-{1\over 4}x'I(x')]-x'(2x+\bar x)R(x,t)\Big\},$$
$$\theta_4(x,t)=|\vec q|^2I_0(x,t)-x'R(x,t).\eqno(3.12)$$
Here the function $I(x')$ is defined as 
\setcounter{equation}{12}
\begin{equation}
I(x')= \left\{ \begin{array}{ll}
 {2\pi\over|\vec q|\chi}{\rm arctan}{2|\vec q|\chi\over x'}, & \chi^2>0\\
   & \\
 {\pi\over|\vec q|p}{\rm log}\Big[\Big(1+{2|\vec q|p\over x'}\Big)\Big/
 \Big(1-{2|\vec q|p\over x'}\Big)\Big], & \chi^2=-p^2<0,\\
   \end{array} \right.
\end{equation}
with $\bar x$ and $x'$ defined as:
$$\bar x=4m_q^2-t-x,\qquad x'={1\over 2}(t-\bar\nu^2).\eqno(3.14)$$
Finally the function $R(x,t)$ in (3.12) is defined as
$$R(x,t)={1\over \bar x}[I(x')-x'I_0(x)].\eqno(3.15)$$

For the definition of the $\pi-$gluon exchange potential
the iterated single pion and gluon exchange interactions
(2.9) have to be subtracted from these expressions
(i.e. (3.5), (3.8)) for the $\pi-$gluon exchange amplitude.\\

\setcounter{section}{4}
\centerline{\bf 4. The Iterated $\pi-$Gluon Exchange
Interaction}

\vspace{0.5cm}

The explicit form for the iterated $\pi-$gluon exchange
interaction in (2.13b) is
$$J={8\pi\over3}f_{\pi qq}^2\alpha_s{m_q^2\over m_\pi^2}
\int{d^2k\over
(2\pi)^3}{\vec\tau^1\cdot\vec\tau^2\over E_k(k^2-q^2-i\epsilon)}$$
$$\times\left\{ {\gamma_5^{(1)}\gamma_5^{(2)}N(\vec k)
\gamma^{(1)}\cdot\gamma^{(2)}\over[m_\pi^2+(\vec p\,'-\vec k)^2]
[m_G^2+(\vec p-\vec k)^2]}
f[(\vec p\,'-\vec k)^2]g[(\vec p-\vec k)^2]\right.$$
$$\left.+{\gamma^{(1)}\cdot\gamma^{(2)}N(\vec k)
\gamma_5^{(1)}\gamma_5^{(2)}\over[m_G^2+(\vec p\,'-\vec k)^2]
[m_\pi^2+(\vec p-\vec k)^2]}
g[(\vec p\,'-\vec k)^2]f[(\vec p-\vec k)^2] \right\} .\eqno(4.1)$$
Here we have used the abbreviations
$$E_k=\sqrt{k^2+m_q^2},$$
$$N(\vec k)=(-\gamma^{(1)}\cdot\vec k-i\gamma_4^{(1)}E_k-im_q)
(\gamma^{(2)}\cdot\vec k-i\gamma_4^{(2)}E_k-im_q).\eqno(4.2)$$

The spin-structure of the integral (4.1) may be
expressed in terms of 5 spin amplitudes $R_j$ defined in
Ref. \cite{ChuR} as 
$$R_1=\gamma_5^{(1)}\gamma_5^{(2)}(\gamma^{(1)}
\cdot\gamma^{(2)}),\quad
R_2=\gamma_4^{(1)}\gamma_4^{(2)} R_1,$$
$$R_3=\{(\gamma_4^{(1)}+\gamma_4^{(2)}),R_1\},\quad
R_4=\gamma^{(1)}\cdot\gamma^{(2)} R_1,
\quad R_5=\gamma_5^{(1)}\gamma_5^{(2)}.\eqno(4.3)$$
The integral (4.1) then takes
the form
$$J=\sum_{i=1}^5j_i(t,p^2)R_i\vec\tau^1\cdot\vec\tau^2,
\eqno(4.4)$$
where the scalar amplitudes $j_i$ may be expressed as dispersion
integrals:
$$j_i(t,p^2)={1\over\pi}\int_{t_0}^\infty dt'
{\bar\eta_i(t',p^2)\over t'-t}.\eqno(4.5)$$
Here the lower integration limit is
$t_0=(m_\pi+m_G)^2$.\\

The spectral functions $\bar\eta_i$ in these integrals have the
explicit expressions

\setcounter{equation}{5}
\begin{equation}
\begin{array}{rcl}
\bar\eta_1(t,p^2)&=&{4m_q^2\beta\over\kappa^4}\{I_1[|\vec q|^2(t-4p^2)
+2\lambda^2]-2\xi I_2+2I_3\},
\\&&\\
\bar\eta_2(t,p^2)&=&\beta\left\{E^2I_1\left[1+{1\over\kappa^2}
\left({\bar\mu^4\over t}+4p^2-{32|\vec q|^2p^2\over\kappa^2}
+{8\lambda^2\over\kappa^2}\right)\right]\right.
\\&&\\
&&\qquad \left.+I_2\left[1+{4E^2\over\kappa^2}
\left(1-{2\xi\over\kappa^2}\right)
\right]+{8E^2I_3\over\kappa^4}-{4E\pi\over\kappa^2}
-{2E(4p^2+\bar\nu^2)\over\kappa^2}I_4\right\},
\\&&\\
\bar\eta_3(t,p^2)&=&{m_q\beta\over2\kappa^2}\left\{EI_1\left(\bar\nu^2
-{\bar\mu^4\over t}+{32|\vec q|^2p^2\over\kappa^2}
-{8\lambda^2\over\kappa^2}\right)\right.
\\&&\\
&&\qquad\qquad \left.+2EI_2\left({4\xi\over\kappa^2}-1\right)
-{8E\over\kappa^2}I_3+2\pi+I_4(\bar\nu^2-t)\right\},
\\&&\\
\bar\eta_4(t,p^2)&=&{\beta\over\kappa^2}\{[4|\vec q|^2p^2-\lambda^2]I_1
+\xi I_2-I_3\},
\\&&\\
\bar\eta_5(t,p^2)&=&\beta\left\{I_1\left
[\delta\left({16|\vec q|^2p^2\gamma
\over\kappa^2}-{4\lambda^2\gamma\over\kappa^2}+
{4p^2\bar\mu^4\over\kappa^2t^2}
-{4p^2\over\kappa^2}\right)-2m_q^2\left({4p^2+\bar\nu^2\over\kappa^2}
-{\bar\mu^2\over t}\right)\right]\right.
\\&&\\
&&\qquad\left.-{I_2\over\kappa^2}[4m_q^2+4\delta(1-\xi\gamma)]
-{4\delta\gamma\over\kappa^2}I_3+{8E\pi\over\kappa^2}
+4EI_4\left[{4p^2+\bar\nu^2\over\kappa^2}-{\bar\mu^2\over t}\right]\right\}
.\\
\end{array}
\end{equation}
Here the notation is
$$\beta=
-{2 f_{\pi qq}^2\alpha_S\over 3\sqrt{t}}
({m_q\over m_\pi})^2,\qquad 
\lambda^2=m_G^2m_\pi^2,\qquad\gamma={2\over\kappa^2}+{1\over t},$$
$$\delta=2p^2+m_q^2,\qquad \xi=t+p^2-\bar\nu^2,\qquad
\bar\mu^2=m_G^2-m_\pi^2,$$
$$\bar\nu^2=m_G^2+m_\pi^2,\qquad \kappa^2=4p^2+t,\qquad
E=\sqrt{p^2+m_q^2}\ .\eqno(4.7)$$
The functions $I_k$ in these expressions are defined as

\setcounter{equation}{7}
\begin{equation}
\begin{array}{rcl}
I_1(t)&=&{1\over E}I_4+
{2\over(p^2-\sigma_-)\sqrt{\sigma_++m_q^2}}\left\{
{\sigma_-+m_q^2\over E^2}\Pi(\gamma,\alpha)-F(\alpha)\right\},\\
&&\\
I_2(t)&=&{2\over\sqrt{\sigma_++m_q^2}}F(\alpha),\\
&&\\
I_3(t)&=&{2\over\sqrt{\sigma_++m^2_q}}\{(\sigma_+ +m_q^2)E(\alpha)
-m_q^2F(\alpha)\},\\
&&\\
I_4(t)&=&{\pi\sqrt{t}\over p}\left\{
{\theta(t_+-t)\over\sqrt{(t_+-t)(t-t_-)}}+
i{\theta(t-t_+)\over\sqrt{(t-t_+)(t-t_-)}}\right\}.\\
\end{array}
\end{equation}
Here $F(\alpha)$, $E(\alpha)$ and $\Pi$ are complete
elliptic integrals of the first, second and third
kind, respectively, and the variables $\sigma_\pm$ are defined as
$$\sigma_\pm={1\over 2}t'+p^2-{1\over 2}\bar\nu^2\pm\kappa
q.\eqno(4.9)$$
The arguments of the elliptic integrals are
$$\alpha={\sigma_+-\sigma_-\over\sigma_++m_q^2},\qquad
\gamma={p^2-\sigma_-\over p^2+m_q^2}.\eqno(4.10)$$
In the expression for $I_4$ the variables $t_\pm$ are defined as
$$t_\pm=t_0+{\lambda^2\over 2p^2}+2\lambda\left\{
\pm\sqrt{1+{\bar\nu^2\over 4p^2}+{\lambda^2\over 16p^4}}-1\right\}.
\eqno(4.11)$$

In order to construct the $\pi-$gluon exchange interaction (2.13b)
the iterated $\pi-$gluon exchange interaction (4.1) has to be 
subtracted from the $\pi-$ gluon exchange amplitude constructed
in section 3 above. For that purpose the latter has to be
expressed in terms of the spin operators $R_j$ (4.3). This
transformation has been given in Ref. \cite{ChuR}. In terms
of these spin operators the $\pi-$gluon exchange amplitude
(3.3) takes the form
$$M=\sum_{j=1}^5r_j(s,t,u)R_j\vec\tau^1\cdot\vec\tau^2.
\eqno(4.12)$$
The scalar amplitudes $r_j$ are linear combinations of the
amplitudes $s_j$ in (3.3). These may also be written as
dispersion integrals:
$$r_j(s,t,u)={1\over\pi}\int_{t_0}^\infty dt'
{\eta_j(t',s,u)\over t'-t}.\eqno(4.13)$$
The spectral functions $\eta_j$ here are the following
linear combinations of the spectral functions $\sigma_j$
in (3.8):
\setcounter{equation}{13}
\begin{equation}
\begin{array}{rll}
\eta_1&=\sigma_1-2m_q^2\sigma_2+2m_q\sigma_3,
&\eta_2=-8E^2\sigma_2+\sigma_4,\\
&&\\
\eta_3&=2Em_q\sigma_2-E\sigma_3,&\eta_4=\sigma_4,\\
&&\\
\eta_5&=(12E^2-2m_q^2)\sigma_2+2m_q\sigma_3+\sigma_5.\\
\end{array}
\end{equation}

Subtraction of the iterated $\pi-$gluon exchange interaction from
the $\pi-$gluon exchange amplitude gives the final
$\pi-$gluon exchange interaction in the form 
$$U_{\pi G}=\sum_{j=1}^5R_j\vec\tau^1\cdot\vec\tau^2
{1\over\pi}\int_{t_0}^\infty dt'
{\eta_j(s,t',u)-\bar\eta_j(t',s)\over t'-t}.\eqno(4.15)$$
Here $s=4m_q^2+4p^2$ in the argument of $\bar\eta_j$.
Constructed in this way the $\pi-$gluon exchange interaction
is real up to the pion production threshold, and only very 
weakly energy dependent.
\vspace{1cm}

\setcounter{section}{5}
\centerline {\bf 5. The $\pi-$Gluon Exchange Potential}

\vspace{0.5cm}

Constituent quarks confined within baryons and with masses
in the 300-400 MeV range have large velocities and call for
a Poincar\'e invariant treatment. The Blankenbecler-Sugar
equation framework admits such a treatment, as demonstrated
e.g. by the calculation of nucleon-nucleon scattering
observables in Ref. \cite{VerW}. In practice this calls
for a helicity amplitude decomposition of the spin
operators $R_j$ (4.3) and solution of the resulting
coupled linear integral equations. This approach, although
straightforward, does not lead to very transparent
results however. For a qualitative estimate of the
numerical strength of the $\pi-$gluon exchange interaction,
and of its importance in comparison to other interaction
components, the standard adiabatic limit representation
in terms of potential operators is most useful.\\

To lowest order in the nonlocal operator 
$(\vec p\,'+\vec p)$ the potential may be expressed in terms
of the potential operators:
$$\tilde\Omega_C=1,\qquad \tilde\Omega_{LS}={1\over 2}i
(\vec\sigma^1+\vec\sigma^2)\cdot\vec p\,'\times\vec p,$$
$$\tilde\Omega_T=\vec\Delta^2(\vec\sigma^1\cdot\vec\sigma^2)
-3(\vec\sigma^1\cdot\vec\Delta)(\vec\sigma^2\cdot\vec\Delta),$$
$$\tilde\Omega_{SS}=\vec\sigma^1\cdot\vec\sigma^2,\eqno(5.1)$$
where $\vec\Delta=\vec p\,'-\vec p$. In the adiabatic limit
the $\pi-$gluon exchange interaction then takes the following
form, once the spin operators $R_j$ are expanded in terms
of the potential operators $\Omega_\alpha$:
$$v(\vec p\,',\vec p)=-\sum_{\alpha=1}^4\sum_{j=1}^5\vec\tau^1\cdot\vec\tau^2
\tilde\Omega_\alpha X_{\alpha j}{1\over\pi}\int_{t_0}^\infty dt'
{\eta_j-\bar\eta_j\over t'-t}.\eqno(5.2)$$
Here $X_{\alpha j}$ is the linear transformation matrix between
the spin operators $R_j$ and the potential operators $\Omega_\alpha$:
\setcounter{equation}{2}
\begin{equation}
X=\left[
\begin{array}{ccccc}
0 & 0 & 0 & \Delta^2/2m_q^2 & 0\\
1/2m_q^2 & -1/2m_q^2 & 0 & -5/2m_q^2 & 0\\
-1/12m_q^2 & 1/6m_q^2 & 0 & 1/3m_q^2 & 1/12m_q^2\\ 
-1 & -1 & -4 & -1 & -\Delta^2/12m_q^2 \\
\end{array} \right].
\end{equation}

For the visualization of the interaction it is useful to
perform a Fourier transformation to configuration space, by which the
potential takes the form 
$$V(r)=\sum_\alpha v_\alpha(r)\Omega_\alpha\vec\tau^1\cdot\vec\tau^2.
\eqno(5.4)$$
Here the spin operators $\Omega_\alpha$ are defined in the
conventional way as
$$\Omega_C=1,\qquad \Omega_{LS}={1\over 2}
(\vec\sigma^1+\vec\sigma^2)\cdot\vec L,$$
$$\Omega_T=3(\vec\sigma^1\cdot\hat r)(\vec\sigma^2\cdot\hat r)
-\vec\sigma^1\cdot\vec\sigma^2,$$
$$\Omega_{SS}=\vec\sigma^1\cdot\vec\sigma^2.\eqno(5.5)$$
For the potential functions $v_\alpha(r)$ the following
explicit expressions obtain:
\setcounter{equation}{5}
\begin{equation}
\begin{array}{rcl}
v_C(r)&=&{1\over 4\pi^2}\int_{t_0}^\infty dt'\sqrt{t'}\eta_C(t')Y_0(r\sqrt{t'}),
\\&&\\
v_{LS}(r)&=&-{1\over 4\pi^2}\int_{t_0}^\infty dt't'\sqrt{t'}\eta_{LS}(t')
Y_0(r\sqrt{t'})\left(1+{1\over r\sqrt{t'}}\right)/r\sqrt{t'},\\
&&\\
v_T(r)&=&{1\over 4\pi^2}\int_{t_0}^\infty dt't'\sqrt{t'}\eta_T(t')
Y_0(r\sqrt{t'})\left(1+{3\over r\sqrt{t'}}+{3\over r^2t'}\right),\\
&&\\
v_{SS}(r)&=&{1\over 4\pi^2}\int_{t_0}^\infty dt'\sqrt{t'}\eta_{SS}(t')
Y_0(r\sqrt{t'}).\\
\end{array}
\end{equation}
Here $Y_0$ is the Yukawa function 
$$Y_0(x)=e^{-x}/x.\eqno(5.7)$$
The ``mass spectrum'' functions $\eta_\alpha$ are defined as
$$\eta_\alpha(t')=-\sum_{j=1}^5X_{\alpha j}[\eta_j(t',s=4m_q^2,u=0)
-\bar\eta_j(p^2=0,t')].\eqno(5.8)$$
Note that the integrals (5.6) are convergent even in the
absence of a pion-quark vertex function. In that case the
ultraviolet divergences in the dispersion integrals (3.5) and
(4.5) may be isolated into delta functions by replacing the
dispersion integrals by suitably subtracted versions.

\vspace{1cm}

\setcounter{section}{6}
\centerline {\bf 6. Numerical Results}

\vspace{0.5cm}

	The controlling parameters for the $\pi-$gluon exchange
interaction are the pion-quark coupling constant $f_{\pi qq}$,
the coupling constant for gluons to constituent quarks
$\alpha_S$, the chiral restoration scale $\Lambda_\chi$ and
the inverse screening length parameter $\Lambda$ in the
one-gluon exchange potential. Of these the pion-quark
coupling constant and the chiral restoration scale parameters
are probably the least contentious. The former may be
related by standard quark model algebra to the 
pseudovector $\pi NN$ coupling constant as
$f_{\pi qq}={3\over 5} f_{\pi NN}$ (see section 2 above).
The value of the latter is $f_{\pi NN}\simeq 1$. For the
chiral restoration scale the typical value is \cite{Mano}
$$\Lambda_\chi\simeq 4\pi f_\pi,\eqno(6.1)$$
where $f_\pi$ is the pion decay constant. With $f_\pi$=93 MeV
one obtains $\Lambda_\chi\simeq$ 1170 MeV. We shall here
use the value $\Lambda_\chi$ = 1000 MeV.\\

The value of the coupling strength of gluons to constituent
quarks $\alpha_S$ is not well established. 
We shall use $\alpha_S=0.42$ in combination with the
screening and fall-off factor (2.12) so as to achieve agreement
with the lattice calculation results reported in Ref. \cite{Lepage}
for momentum transfer values that correspond to the charmonium
and bottomonium mass scales. As pointed out in section 2 
this implies a very weak quark-gluon coupling for
momenta that correspond to the light constituent quark
mass scale (i.e. $\alpha_S\simeq 0.04-0.27$, depending on
the choice of screening parameter $\Lambda$).
Careful numerical studies \cite{Papp} of the baryon
spectrum with hyperfine interaction models that are
linear combinations of the single
gluon and pion exchange interactions \cite{Vento, Miller}
indicate that 
$\alpha_S$ has to be less than about 0.35 if the 
problem of incorrect ordering of the lowest positive
and negative parity states is to be avoided. \\

For the inverse screening length parameter $\Lambda$ we
shall use both the chiral restoration  scale and the
confinement scale $\Lambda_{QCD}$. As the latter is 
much smaller ($\simeq$ 250 MeV \cite{Don}), the resulting 
$\pi-$gluon exchange interaction, with the exception of the 
spin-spin component, will be stronger for the latter value. 
The qualitative features will be the
same however. There is obviously a possibility of
balancing the values of $\alpha_S$ and the screening
parameter against one another.\\

In Figs. 2 and 3 the central and spin-orbit components
of the $\pi-$gluon exchange interaction are shown as functions of
quark separation $r$, along with the corresponding Fourier transform of 
the components of the single gluon exchange interaction
(note that the latter lack the flavor factor $\vec\tau_1
\cdot\vec\tau_2$). These components of the $\pi-$gluon
exchange interaction have very short range ($\le 0.3$ fm),
and are weak. The most notable feature is the smallness
of the spin-orbit component of the $\pi-$gluon exchange 
interaction. This is a phenomenologically desirable feature
as the spin-orbit splitting of the $P$-shell of the
baryon spectrum is very small. The smallness of the spin-orbit
component is  a consequence of the fact that the ``mass spectrum''
function $\eta_{LS}$ in the spin-orbit potential $v_{LS}(r)$
(5.6) has a zero.\\ 

In Figs. 4 and 5 the tensor and spin-spin components of the
$\pi-$gluon exchange interaction are given as functions of
$r$. For comparison the corresponding components of the Fourier
transform of the one-pion-exchange potential for quarks are also 
shown in Fig. 5.
The most notable result here is that the strength of tensor
component of the $\pi-$gluon exchange interaction is of the
same order of magnitude as that of the one-pion-exchange
interaction, although it has opposite sign. As a consequence
the tensor component of the $\pi-$gluon exchange interaction
tends to cancel that of the one-pion-exchange interaction.
This cancellation of the main flavor dependent tensor interaction
between the quarks provides a simple explanation for why
the hyperfine splitting of the baryon spectrum is so well 
described by a simple pion exchange interaction, provided
that its tensor component is dropped \cite{Pless}.\\

In the shortest range region the $\pi-$gluon
tensor interaction is the predominant one, which is
phenomenologically desirable, as the pion-exchange tensor
by itself would split the spin-symmetric negative
parity multiplet $N(1650)-N(1700)-N(1675)$ incorrectly
\cite{GloRis}.\\

The spin-spin component of the $\pi-$gluon exchange interaction
has the same sign, and is of similar magnitude as the
corresponding component of the one-pion-exchange 
interaction. This suggests that it is a significant component
of the hyperfine interaction between quarks, even in the
case of a very weak coupling between gluons and
constituent quarks. This also contributes to the
understanding of the phenomenological fact that
the hyperfine splitting of the baryon spectrum can
be explained by a strong attractive flavor dependent
spin-spin interaction \cite{GloRis,Dann}.\\

The sign of the spin-spin component of the $\pi-$gluon
exchange interaction changes at very short range. This
sign change is a consequence of the weakening
at high momentum transfer of the effective quark-gluon
coupling constant (2.11), (2.12). As shown in Fig. 5
this feature is lost if the inverse screening length
parameter $\Lambda$ is set to the chiral restoration scale
$\Lambda_\chi$ (1 GeV).\\ 

\vspace{1cm}

\setcounter{section}{7}
\centerline {\bf 7. Discussion}

\vspace{0.5cm}

The main result of the present investigation is that even
in the case of very weak coupling of gluons to
constituent quarks, the irreducible $\pi-$ gluon exchange
interaction is of significant magnitude. The reason for this
is that even if the gluon exchange is screened for long
wavelengths, the $\pi-$gluon exchange loop diagrams allow
exchange of short wavelength gluons, along with long wavelength
pions. The result indicates that the interaction between
constituent quarks should be far more intricate than the conventional
single gluon exchange and (less conventional for quarks) one-pion-
exchange interactions. While the latter will be screened out
at short range, the $\pi-$gluon exchange interaction is not
restricted in that way.\\

The results suggest that the hyperfine interaction between
light constituent quarks should be formed of a (weak)
single gluon exchange component, a pion exchange component
with conventional strength and an irreducible $\pi-$gluon
exchange interaction. The single gluon exchange interaction
is weak because the effective quark-gluon coupling is
weak in the infrared limit \cite{Matting,Brodsky}, and
therefore the problem of a large gluonic spin-orbit
interaction \cite{Carls} is avoided. The pion exchange
tensor interaction is in effect cancelled by the large
tensor component of the $\pi-$gluon exchange interaction,
and thus the incorrect (small) spin-orbit splitting of the
low lying negative parity resonances is avoided. The
$\pi-$gluon exchange and single pion exchange interactions
combine to a strong attractive flavor dependent spin-spin
interaction, which brings the low lying positive
parity resonances below the lowest negative parity resonances
in agreement with experiment.\\

There are a large number of other exchange mechanisms that may 
contribute significantly to the hyperfine interaction between
quarks. Among these are vector meson exchange \cite{GloRis} 
and two-pion exchange, which definitely
merit further study. The two-pion
exchange interaction mainly leads to a flavor independent
attractive interaction, which contributes to the strength
of the effective confining interaction at short range. There
will be a strong overlap between that interaction and the
vector meson exchange interactions.\\

The presence of an irreducible $\pi-$gluon exchange interaction
is an immediate consequence if there exists a pion exchange
and (however screened) gluon exchange interaction between
quarks. It is also independent of the structure of the
pion itself -- whether it arises as a succession of
instanton induced quark-antiquark interactions, according to
a common view, or whether is has a simpler quark-antiquark
structure.\\ 

There are several reasons to believe that the single gluon
exchange interaction between constituent quarks should be very
weak beyond the chiral restoration scale or at least beyond
the confinement scale. Among these are the absence of gluonic
degrees of freedom in cooled lattice calculations \cite{Nege}
and the very weak residual gluonic coupling in the valence-QCD
approximation \cite{KFLiu}. 
Phenomenological investigations of the behavior of the
running coupling strength of quarks to gluons in the
infrared limit also indicate that the coupling constant
drops or freezes at a constant value \cite{Matting,Brodsky}.
The (possible) disconnection of gluons from
the constituent quarks at long range may be described
phenomenologically in several ways. The most direct is the
introduction of a suitable range dependence in the effective
coupling constant $\alpha_S$, which at very long range 
-- i.e. beyond the confinement scale --  weakens
the coupling constant, which is known to increase with range
in the short range region. In practice this would bring
about the screening effect described here simply by the
form (2.11). \\

The substantial size of the $\pi-$gluon 
tensor and spin-spin interaction components
suggests that the interaction between constituent quarks
in the end may prove to be as complex as the nucleon-nucleon
interaction proved to be, and that it -- at least partly --
has to be constructed phenomenologically as is the case
with the latter \cite{Tak,Bali}.\\

\vspace{1cm}

\centerline {\bf Acknowledgments}

\vspace{0.5cm}

D. O. R. thanks Professor S. J. Brodsky for helpful
correspondence. This work has been supported in part 
by the Academy of Finland under contract 34081. 

\vspace{1cm}

\centerline{\bf Figure Captions}

\vspace{0.5cm}

Figure 1. Feynman diagram representation of the
$\pi-$gluon exchange quark-quark scattering amplitude.\\

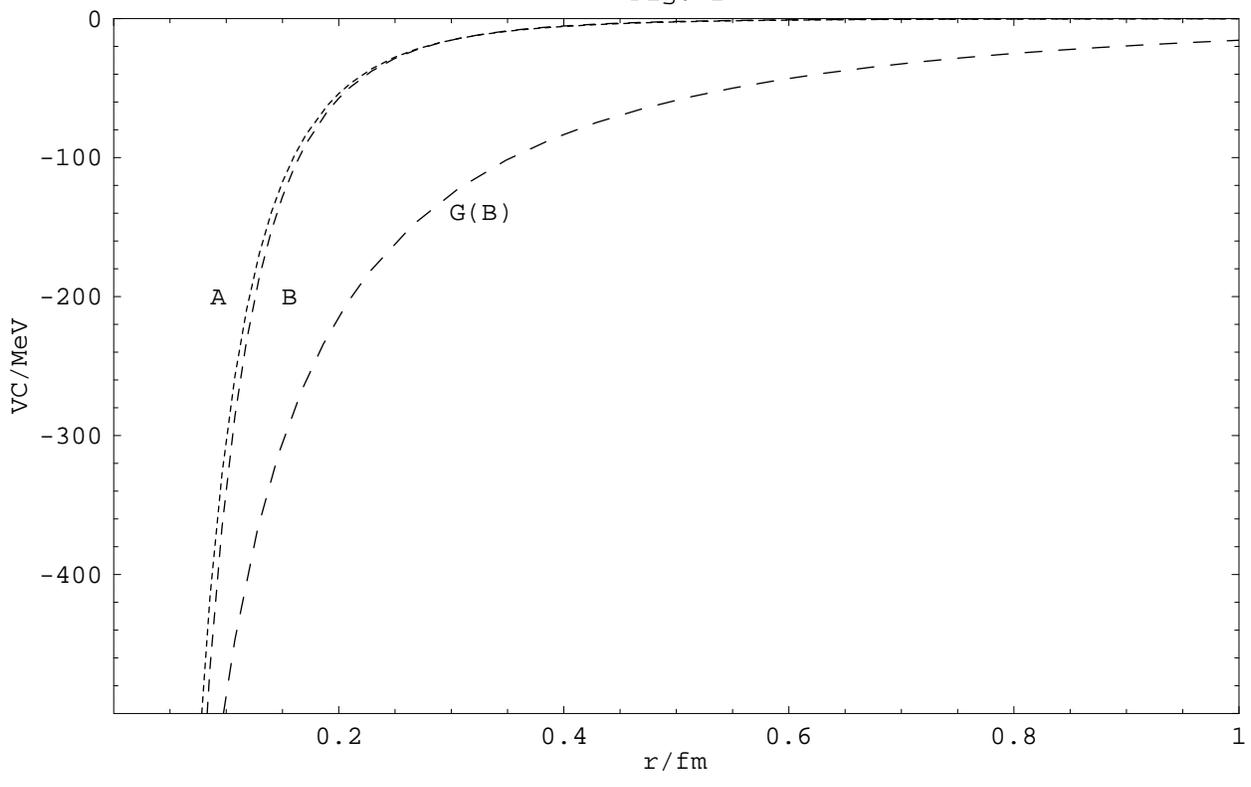
Figure 2. The central component of the 
$\pi-$gluon exchange potential as function of 
quark separation. The curve A is the result when the
screening function g($k^2$)
in the gluon exchange interaction is taken to be simply the
fall-off factor $1/(1+{k^2\over\Lambda^2_F})$ and the curve B
is the result when the full form (2.12) for the screening
function is used with the screening parameter $\Lambda$
equal to $\Lambda_{QCD}$ (250 MeV). The curve G(B) is
the corresponding single gluon exchange interaction without
the fall-off factor.
\\

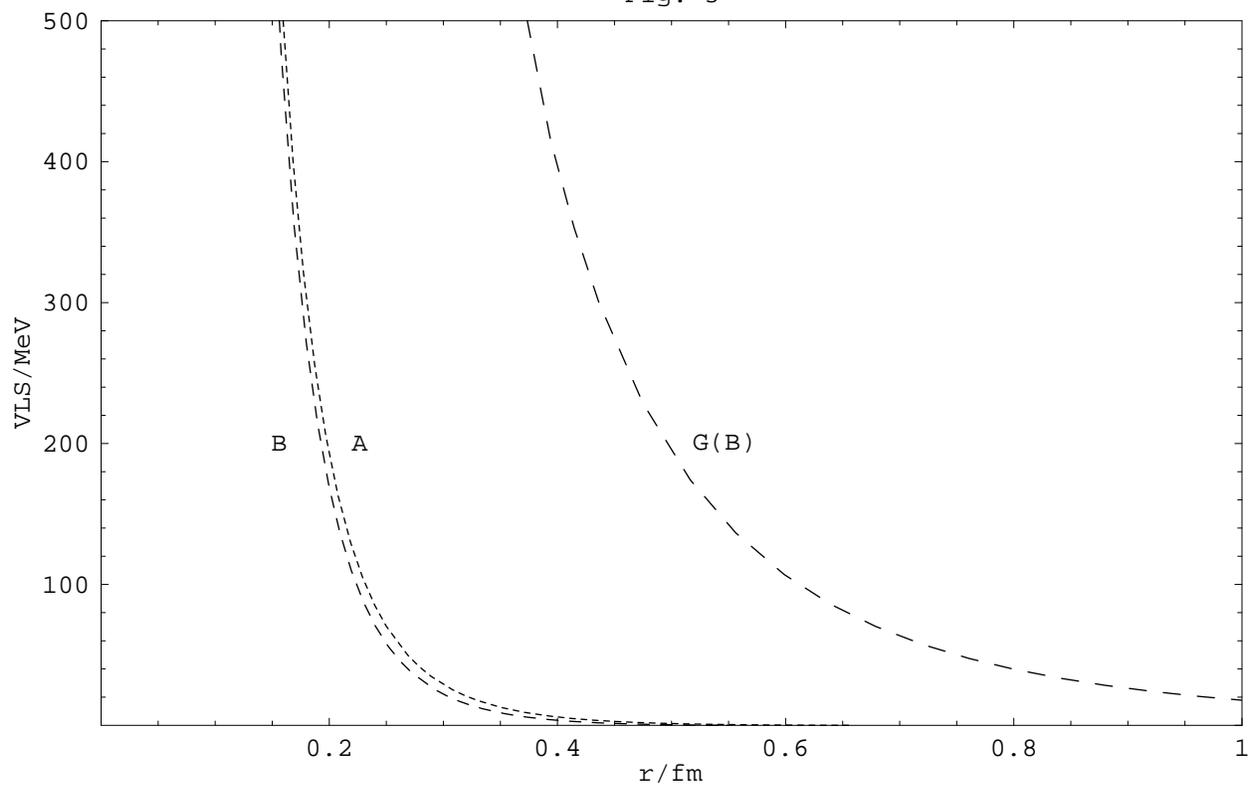
Figure 3. The spin-orbit component of the 
$\pi-$gluon exchange potential as function of 
quark separation. The curve A is the result when the
screening function g($k^2$)
in the gluon exchange interaction is taken to be simply the
fall-off factor $1/(1+{k^2\over\Lambda^2_F})$ and the curve B
is the result when the full form (2.12) for the screening
function is used with the screening parameter $\Lambda$
equal to $\Lambda_{QCD}$ (250 MeV). The curve G(B) is
the corresponding single gluon exchange interaction without
the fall-off factor.
\\

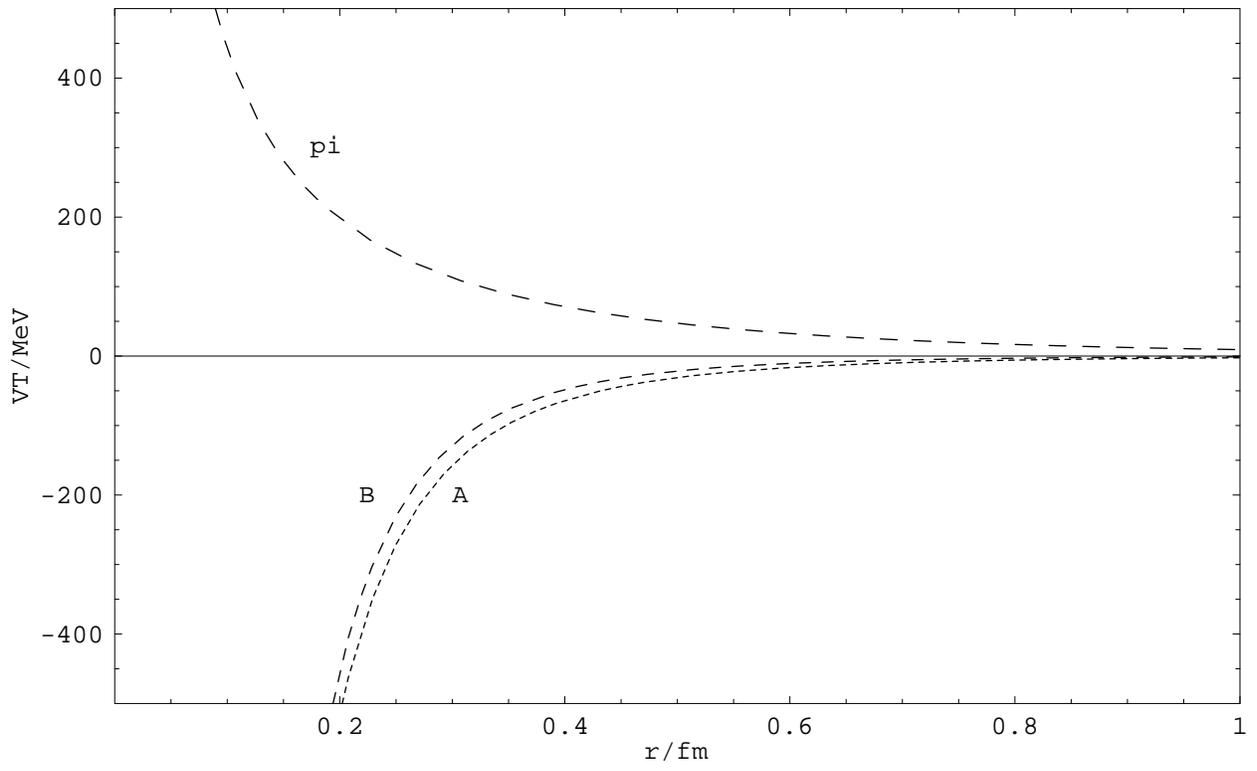
Figure 4. The tensor component of the 
$\pi-$gluon exchange potential as function of 
quark separation. The curve A is the result when the
screening function g($k^2$)
in the gluon exchange interaction is taken to be simply the
fall-off factor $1/(1+{k^2\over\Lambda^2_F})$ and the curve B
is the result when the full form (2.12) for the screening
function is used with the screening parameter $\Lambda$
equal to $\Lambda_{QCD}$ (250 MeV). The curve pi is
the corresponding one-pion exchange potential.
\\

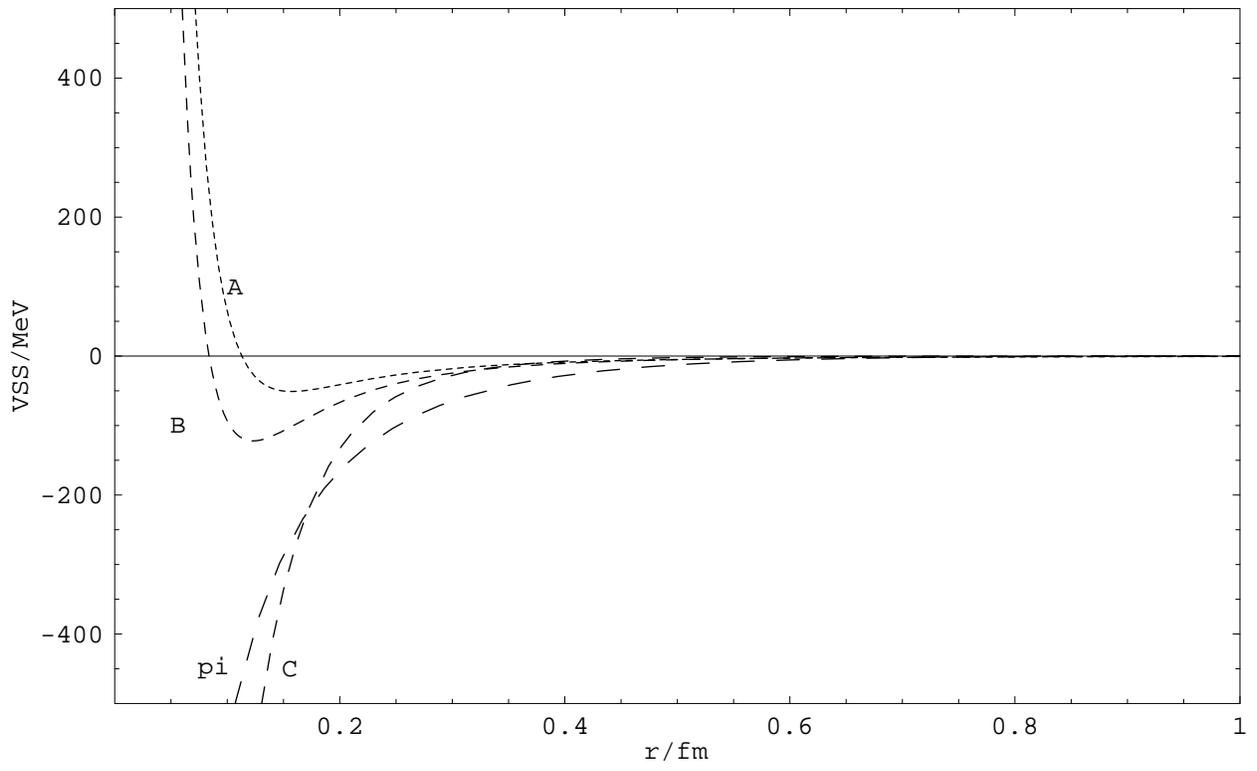
Figure 5. The spin-spin component of the 
$\pi-$gluon exchange potential as function of 
quark separation. The curve A is the result when the
screening function g($k^2$)
in the gluon exchange interaction is taken to be simply the
fall-off factor $1/(1+{k^2\over\Lambda^2_F})$ and the curves B
and C are the results when the full form (2.12) for the screening
function is used with the screening parameter $\Lambda$
equal to $\Lambda_{QCD}$ (250 MeV) and $\Lambda_\chi$ (1 GeV),
respectively. The curve pi is
the corresponding one-pion exchange potential.
\\

\end {document}
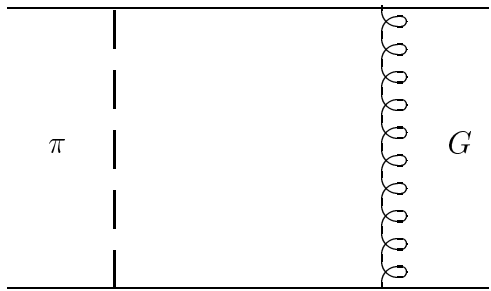
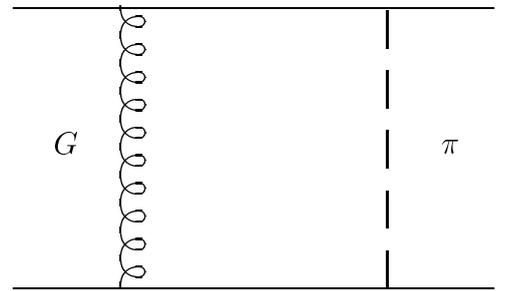
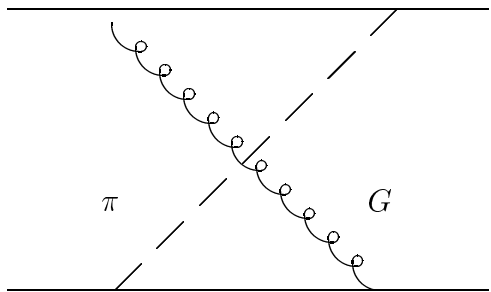
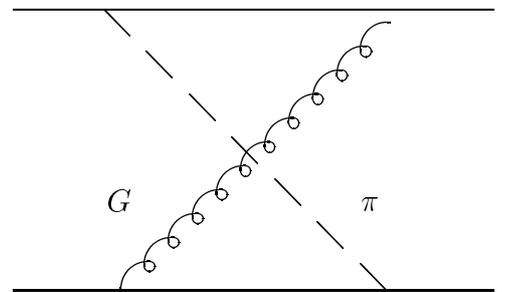
